\documentclass{eas}
\usepackage{graphicx}
\usepackage{amssymb}
\begin{document}

\title{The Deuteration Clock for\\Massive Starless Cores} 
\author{Shuo Kong}\address{Dept. of Astronomy, University of Florida, Gainesville, Florida 32611, USA}
\author{Jonathan C. Tan}\address{Depts. of Astronomy \& Physics, University of Florida, Gainesville, Florida 32611, USA}
\author{Paola Caselli}\address{Max-Planck-Institue for Extraterrestrial Physics, D-85748 Garching, Germany}
\author{Francesco Fontani}\address{INAF - Osservatorio AstroÞsico di Arcetri, I-50125, Florence, Italy}
\begin{abstract}
To understand massive star formation requires study of its initial
conditions.
Two massive starless core candidates, C1-N \& C1-S, have been detected
in IRDC G028.37+00.07 in $\rm N_2D^+$(3-2) with {\it ALMA}. From their
line widths, either the cores are subvirial and are thus young
structures on the verge of near free-fall collapse, or they are
threaded by $\sim1$~mG $B$-fields that help support them in near
virial equilibrium and potentially have older ages.
We modeled the deuteration rate of $\rm N_2H^+$ to constrain collapse
rates of the cores.  First, to measure their current deuterium fraction, $D_{\rm
  frac}^{\rm N_2H^+}$ $\equiv [\rm N_2D^+]/[N_2H^+]$, we observed
multiple transitions of $\rm N_2H^+$ and $\rm N_2D^+$ with {\it
  CARMA}, {\it SMA}, {\it JCMT}, {\it NRO 45m} and {\it IRAM 30m}, to
complement the {\it ALMA} data. For both cores we derived $D_{\rm
  frac}^{\rm N_2H^+}\sim0.3$, several orders of magnitude above the
cosmic [D]/[H] ratio.
We then carried out chemodynamical modeling, exploring how 
collapse rate relative to free-fall, $\alpha_{\rm ff}$, affects the
level of $D_{\rm frac}^{\rm N_2H^+}$ that is achieved from a given
initial condition. To reach the observed $D_{\rm frac}^{\rm
  N_2H^+}$, most models require slow collapse with $\alpha_{\rm
  ff}\sim0.1$, i.e., $\sim1/10$th of free-fall.
This makes it more likely that the cores have been able to reach a
near virial equilibrium state and we predict that strong $B$-fields
will eventually be detected. The methods developed here will be useful
for measurement of the pre-stellar core mass function.
\end{abstract}
\maketitle
\section{Deuteration as a Tracer of Massive Starless Core Dynamics}

Massive star formation involves many open questions, in part because
examples of initial conditions of the process are rare, distant and
deeply embedded in massive gas clumps, i.e., structures that
eventually fragment into star clusters (Tan et al. 2014). A key
question is whether the formation mechanism is a scaled-up version of
low-mass star formation. Core Accretion models assume it is: e.g., the
Turbulent Core model (McKee \& Tan 2003 [MT03]) adopts an initial
condition of a virialized massive starless core---a coherent gas
cloud that will collapse via a central disk to form a single star
or small-$N$ multiple. Alternatively, the Competitive Accretion model
(Bonnell et al. 2001) forms massive stars at the center of a clump
undergoing rapid global collapse and fragmentation mostly into a swarm
of low-mass stars. A way to test the models is to identify and
characterize initial conditions: do massive starless cores exist and,
if so, what is their dynamical state?

However, a difficulty in such studies is estimating the strength
of $B$-fields, which may provide significant support in addition to
turbulence (note, thermal pressure is unimportant in these massive,
cold clouds). While strong $B$-fields are seen around massive
protostellar cores
(e.g., Zhang et al. 2014), there are few measurements at earlier
stages. Recently, Pillai et al. (2015) measured dynamically strong,
$\sim 1$~mG $B$-fields in dark, presumably starless regions of
Infrared Dark Clouds (IRDCs).

As a complementary approach, we attempt to measure core ages by
astrochemical means, especially the deuteration fraction of key
species, and compare these with dynamical (i.e., free-fall)
timescales. If the chemical age is greater than the free-fall time,
then we expect that a core is likely to have reached approximate
virial equilibrium, so that any global contraction is at a relatively
slow rate, perhaps regulated by $B$-field support. We also couple the
chemical network to simple dynamical models to constrain the collapse rate
relative to free-fall collapse.

The particular indicator we model is the deuterium fraction of $\rm
N_2H^+$, i.e., $D_{\rm frac}^{\rm N_2H^+} \equiv {\rm
  [N_2D^+]/[N_2H^+]}$. It rises in cold, dense conditions of starless
cores by $\gtrsim$4 orders of magnitude above cosmic [D]/[H]
$\sim10^{-5}$.  This is due to the parent exothermic reaction p-$\rm
H_3^+ + HD \rightleftharpoons$ p-$\rm H_2D^+ +$ p-$\rm H_2 + 232\:K$
being favored at low temperatures ($\lesssim30\:$K) (Pagani et
al. 1992). $D_{\rm frac}^{\rm N_2H^+}$ has been observed to be a good
tracer of both low- and high-mass starless cores (e.g., Crapsi et
al. 2005; Tan et al. 2013 [T13]), and is better than
$D_{\rm frac}^{\rm HNC}$ and $D_{\rm frac}^{\rm NH_3}$
(Fontani et al. 2015).  
Other methods, such as dust continuum (e.g., Rathborne et al. 2006),
dust extinction (e.g., Butler \& Tan 2012), are likely subject to
contamination from the much more massive clump envelope surrounding
the cores. Dust continuum emission is also more sensitive to warmer,
protostellar cores, rather than starless cores.

\section{Results for the Massive Starless Cores C1-N and C1-S}

Two massive starless core candidates, C1-N and C1-S, were identified
in IRDC G028.37+00.07, hereafter IRDC C, from their $\rm N_2D^+$(3-2)
emission observed with {\it ALMA} by T13. Based on 1.3~mm dust
continuum emission, C1-S is more massive, with $\sim60\:M_\odot$,
while C1-N has $\sim20\:M_\odot$. The cores are $\sim$0.1~pc in
radius. C1-S is round and centrally concentrated (in both mm continuum
and $\rm N_2D^+$(3-2) integrated intensity), while C1-N appears more
fragmented. Both cores are dark at wavelengths up to $100\:{\rm \mu
  m}$ and there is no sign of star formation activity (e.g., from
SiO(5-4)). Velocity dispersions inferred from $\rm N_2D^+$(3-2) line
widths are about half that of the fiducial MT03 Turbulent Core model,
which assumes large-scale $B$-fields that imply an
Alfv\'en Mach number of $m_A=1$. To achieve virial equilibrium would
require stronger, $\sim 1$~mG fields, which imply $m_A\simeq 0.3$.

Follow-up observations of multiple transitions of $\rm N_2H^+$(1-0),
(3-2), (4-3) and $\rm N_2D^+$(1-0) and (2-1) have been presented by
Kong et al. (2015a [K15a]). Excitation temperatures of $\sim 4$--7~K
are derived for the $\rm N_2D^+$ line emission, resulting in a range
of values of $D_{\rm frac}^{\rm N_2H^+}\simeq0.15$--0.72 for C1-S and
$\simeq0.16$--0.44 for C1-N. However, additional factor of 2
uncertainties arise due to the difficulty in separating the $\rm
N_2H^+$ emission of the cores from their more extended clump
envelopes.


A gas-phase, spin-state astrochemical network was developed by Kong et
al. (2015b [K15b])
to follow time evolution of $D_{\rm frac}^{\rm N_2H^+}$ for given
density ($n_{\rm H}$), temperature ($T$), cosmic ray ionization rate
($\zeta$), extinction of Galactic FUV radiation ($A_V$) and for either
fixed or time-dependent heavy element gas phase depletion factor
($f_D$) via freeze-out onto dust grain ice mantles. For fiducial
conditions of $T\lesssim 15\:$K, $\zeta=2.5\times10^{-17}\:{\rm
  s}^{-1}$, $A_V=30\:$mag 
and $f_D=10$ (i.e., gas phase underabundance by a factor of
10), the K15a results for chemical evolution at the densities of C1-S
($n_{\rm H}\simeq6\times10^5\:{\rm cm^{-3}}$) and C1-N ($n_{\rm
  H}\simeq2\times10^5\:{\rm cm^{-3}}$) are that it takes $\gtrsim
1$~Myr to reach the equilibrium value of $D_{\rm frac,eq}^{\rm
  N_2H^+}\simeq 0.2$--0.3, if starting from an initial OPR$^{\rm
  H_2}$=3. Starting from OPR$^{\rm H_2}$=0.01 only leads to a factor
of 2 reduction in this timescale. Note, the free-fall times for C1-S
and C1-N are (0.5 and 1.0)$\times10^5$~yr, respectively. Thus the
cores are consistent with being in chemical equilibrium and this would
have taken $\gtrsim10$ local free-fall times to achieve.

K15a also present simple chemodynamical modeling of collapsing cores
by controlling the density increase via ${\rm d}n_{\rm H}(t)/{\rm d}t
= \alpha_{\rm ff} n_{\rm H}(t)/t_{\rm ff}(t)$, where $t_{\rm ff}$ is
the local free-fall time at current density $n_{\rm H}$, and
$\alpha_{\rm ff}$ is a dimensionless parameter setting the collapse
rate relative to free-fall. These models are then also parameterized
by a target final density, $n_{\rm H,1}$, an initial density, $n_{\rm
  H,0}$, an initial heavy element depletion factor, $f_{D,0}$, and an
initial OPR$^{\rm H_2}_0$, with $T$, $\zeta$ and $A_V$ being held
fixed.

\begin{figure}[htb!]
\vspace{-0.25in}
\includegraphics[width=12cm]{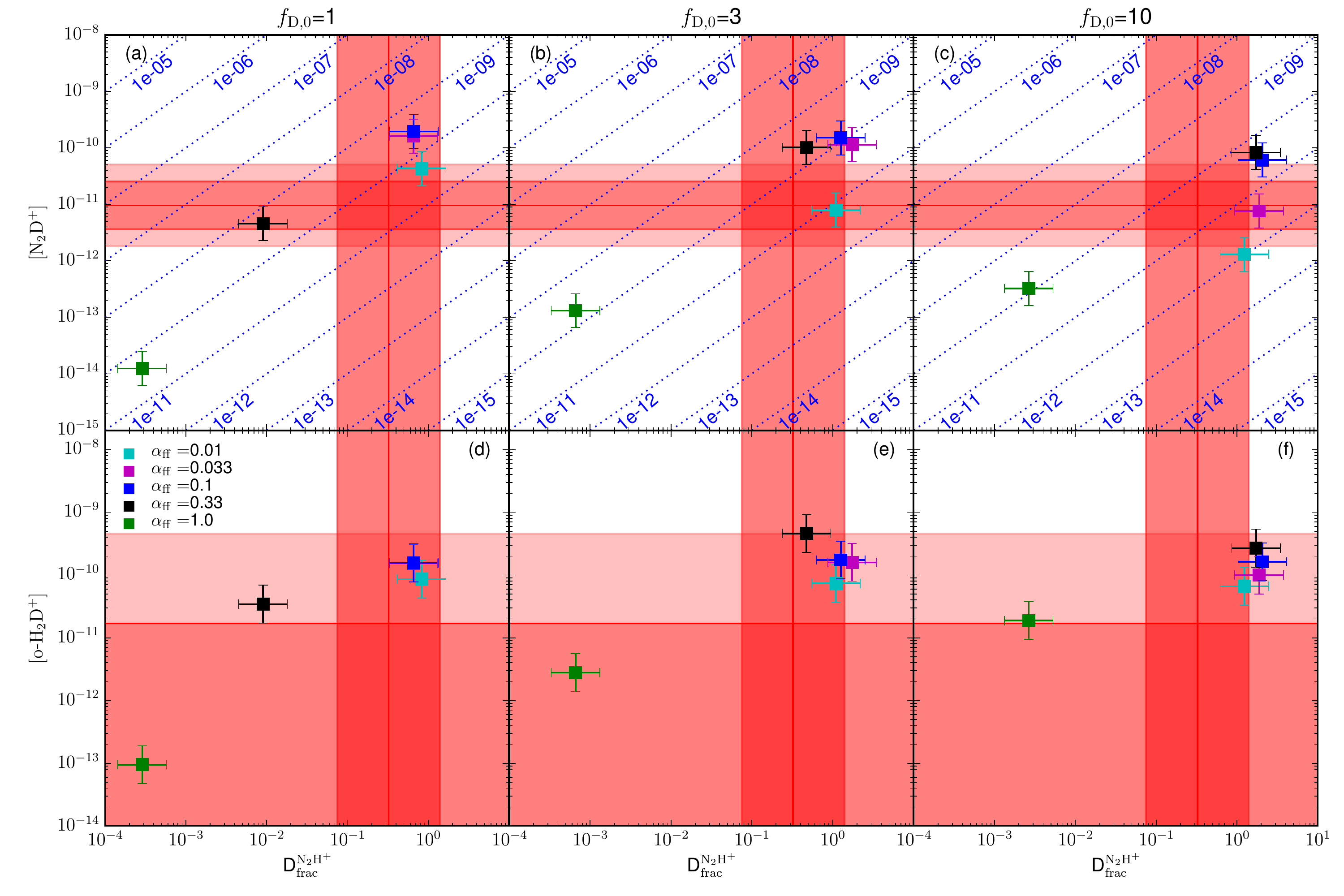}
\vspace{-0.23in}
\caption{
%
(From K15b) C1-S fiducial models ($T=10$~K, $A_V=30$~mag,
  $\zeta=2.5\times10^{-17}\:{\rm s}^{-1}$, OPR$^{\rm H_2}_0=1$,
  $n_{\rm H,0}/n_{\rm H,1}=0.1$ with $f_{D,0}=1,3,10$ [left, middle,
    right columns]) and observational constraints in the [$\rm
    N_2D^+$] - $D_{\rm frac}^{\rm N_2H^+}$ (top row) and [$\rm o$-$\rm
    H_2D^+$] - $D_{\rm frac}^{\rm N_2H^+}$ (bottom row) parameter
  space. The blue dotted lines in the top row show constant [$\rm
    N_2H^+$] values. The red shaded areas show the observational
  constraints.
\label{fig:result}}
\end{figure}

K15b applied these chemodynamical models to the specific cases of C1-S
and C1-N, with $T=10\:$K, $A_V=30\:$mag, $\zeta=$0.1 to
10$\times10^{-17}\:{\rm s}^{-1}$, OPR$^{\rm H_2}_0=$0.01 to 3,
$f_{D,0}=1,3,10$, $n_{\rm H,1}$ set equal to present-day observed
densities, $n_{\rm H,0}$ set to 10 or 100 times smaller, and
$\alpha_{\rm ff}$ explored from 0.01 to 1. Fig.~1 shows example
results for $\alpha_{\rm ff}=0.01,0.033,0.1,0.33,1$ and
$f_{D,0}=1,3,10$ applied to C1-S in the parameter space of present-day
[$\rm N_2D^+$], $D_{\rm frac}^{\rm N_2H^+}$ and [$\rm o$-$\rm
  H_2D^+$].
The observed properties of C1-S, including an upper limit on [$\rm
  o$-$\rm H_2D^+$], also observed by K15a) are shown by red shaded
regions. Models in or near the overlapped red areas are consistent
with all constraints. $D_{\rm frac}^{\rm N_2H^+}$ is the most
stringent constraint and its high observed values of $\sim0.3$ rules
out fast collapsing $\alpha_{\rm ff}=1$ models, regardless of initial
depletion factor.
Models with $\alpha_{\rm ff}\lesssim0.3$ give a much better match. In
fact, the observations are broadly consistent with chemical
equilibrium values, which the slow-collapsing models have time to
converge to. From the broader parameter space exploration of K15b, it
is concluded that 
the most likely evolutionary history of C1-N and C1-S involves
collapse with $\alpha_{\rm ff}\lesssim0.1$.

The above methods and results demonstrate the utility of using $\rm
N_2D^+$ as a tracer of pre-stellar cores, especially in massive clump
environments that exhibit extended $\rm N_2H^+$ emission. The ability
to also estimate a deuteration age will also be helpful for assessing
an unbiased pre-stellar core mass function, i.e., enabling accounting
for the varying lifetimes of the cores.


\begin{thebibliography}{99}
\bibitem[Bonnell et al.(2001)]{2001MNRAS.323..785B} Bonnell, I.~A., Bate, 
M.~R., Clarke, C.~J., \& Pringle, J.~E.\ 2001, MNRAS, 323, 785
\bibitem[Butler 
\& Tan(2012)]{2012ApJ...754....5B} Butler, M.~J., \& Tan, J.~C.\ 2012, ApJ, 754, 5
\bibitem[Crapsi et al.(2005)]{}
Crapsi, A., Caselli, P., Walmsley, M. C., et al. 2005, ApJ, 619, 379
\bibitem[Fontani et 
al.(2015)]{2015A&A...575A..87F} Fontani, F., Busquet, G., Palau, A., et al.\ 2015, A\&A, 575, A87
\bibitem[Kong et al.(2015b)]{2015ApJ...804...98K} Kong, S., Caselli, P., 
Tan, J.~C., Wakelam, V., \& Sipil{\"a}, O.\ 2015b, ApJ, 804, 98 [K15b]
\bibitem[Kong et al.(2015a)]{2015arXiv150908684K} Kong, S., Tan, J.~C., 
Caselli, P., et al.\ 2015a, arXiv:1509.08684 [K15a]
\bibitem[McKee 
\& Tan(2003)]{2003ApJ...585..850M} McKee, C.~F., \& Tan, J.~C.\ 2003, ApJ, 585, 850 [MT03]
\bibitem[Pagani et 
al.(1992)]{1992A&A...258..479P} Pagani, L., Salez, M., \& Wannier, P.~G.\ 1992, A\&A, 258, 479
\bibitem[Pillai et al.(2015)]{2015ApJ...799...74P} Pillai, T., Kauffmann, 
J., Tan, J.~C., et al.\ 2015, ApJ, 799, 74
\bibitem[Rathborne et al.(2006)]{2006ApJ...641..389R} Rathborne, J.~M., 
Jackson, J.~M., \& Simon, R.\ 2006, ApJ, 641, 389
\bibitem[Tan et al.(2014)]{2014prpl.conf..149T} Tan, J.~C., Beltr{\'a}n, 
M.~T., Caselli, P., et al.\ 2014, Protostars and Planets VI, 149
\bibitem[Tan et al.(2013)]{2013ApJ...779...96T} Tan, J.~C., Kong, S., 
Butler, M.~J., Caselli, P., \& Fontani, F.\ 2013, ApJ, 779, 96 [T13]
\bibitem[Zhang et al.(2014)]{} Zhang, Q., Qiu, K., Girart, J. M. et al. 2014, ApJ, 792, 116
\end{thebibliography}
\end{document}